\documentclass[a4paper,11pt]{article}

 \usepackage{float}

\usepackage{jheppub}

\usepackage{amsfonts,graphics,epsfig,subfigure}

\title{Effects of Lovelock gravity on the Joule-Thomson expansion}

\author[a,b,1]{Jie-Xiong Mo,\note{Corresponding author}}
\author[a,b]{Gu-Qiang Li}

 \affiliation[a]{Institute of Theoretical Physics, Lingnan Normal University, Zhanjiang, 524048, Guangdong, China}
\affiliation[b]{Department of Physics, Lingnan Normal University, Zhanjiang, 524048, Guangdong, China}

\emailAdd{mojiexiong@gmail.com}
\emailAdd{zsgqli@hotmail.com}

\abstract{Effects of Lovelock gravity on the Joule-Thomson expansion are probed from various perspectives. The well-known Joule-Thomson coefficient is derived with both the explicit expression and intuitive image presented. Moreover, the inversion curves showing the relation between the inversion temperature and the inversion pressure are studied. It is shown that for given inversion pressure, the inversion temperature of the case $\alpha\neq0$ ($\alpha$ is the Lovelock parameter) is much lower than that of the case $\alpha=0$. And the inversion temperature tends to decrease with $\alpha$, in contrast to the effect of the electric charge. It is also shown that the ratio between the minimum inversion temperature and the critical temperature decreases with $\alpha$ for $\alpha\neq0$. Furthermore, the isenthalpic curves are investigated with rich physics revealed. The intersection point between the isenthalpic curve and the inversion curve is exactly the inversion point discriminating the heating process from cooling process. It is shown that both the inversion temperature and the inversion pressure for $\alpha\neq0$ are much lower for the same given mass of the black hole, showing the effect of Lovelock gravity. Last but not the least, we discuss the case of uncharged Lovelock AdS black holes with interesting feature found. It is shown that the Joule-Thomson coefficient is always positive, suggesting the expansion is always in the regime of cooling process. And no inversion temperature exists, in contrast to the case $Q\neq0$. Isenthalpic curves are also quite different since the temperature increases monotonically with the pressure when the mass is specified.}

\begin{document}
\maketitle
\flushbottom

\section{Introduction}

    Thermodynamics of AdS (Anti-de Sitter) black holes is such an exciting topic that has attracted lots of attention. It was disclosed that there exists phase transition between the Schwarzschild AdS black hole and the thermal AdS space~\cite{Hawking2}. And this transition was then named after its discoverer, Hawking and Page. Concerning charged AdS black holes, rich phase structures were found by Chamblin et al \cite{Chamblin1,Chamblin2}. It was shown that charged AdS black holes are closely related to the liquid-gas system. This observation was further supported by the study of $P-V$ criticality \cite{Kubiznak}. Moreover, it was suggested recently that AdS black holes share many similarities with the ordinary thermodynamic systems in critical behavior \cite{Gunasekaran,Altamirano,Sherkatghanad}. One can refer to the most recent review \cite{Kubiznak2} for details and more references. Apart from the critical phenomena, many other interesting aspects of AdS black holes have also been covered in the recent literatures. For example, the novel concept of holographic heat engine was proposed by Johnson \cite{Johnson1}, suggesting one more way to extract mechanical work from black holes. This idea was further generalized with the proposal that one can even implement black holes as efficient power plant \cite{weishaowen}.

    Recently, Joule-Thomson expansion, a well-known process in classical thermodynamics, was generalized to AdS black holes by \"{o}kc\"{u} and Aydiner \cite{Aydiner1} creatively. In classical thermodynamics, Joule-Thomson expansion refers to the expansion of gas through a porous plug from a high pressure section to a low pressure section via an isenthalpic process. For RN-AdS black holes \cite{Aydiner1} and Kerr-AdS black holes \cite{Aydiner2}, both the inversion curves and isenthalpic curves were investigated. These pioneer works were generalized to quintessence RN-AdS Black Holes \cite{Ghaffarnejad}, holographic superfluids \cite{Yogendran} and RN-AdS Black Holes in $f(R)$ gravity \cite{Chabab}. More recently, we probed in detailed the effect of the dimensionality on the Joule-Thomson expansion \cite{jiexiong3}. It was shown that \cite{jiexiong3} the ratio between minimum inversion temperature and the critical temperature decreases with the dimensionality $d$ while it recovers the result in Ref. \cite{Aydiner1} when $d=4$.

    We are curious to examine our former result in another high-dimensional space-time. On the other hand, we are also interested in the effect of Lovelock gravity on the Joule-Thomson expansion. So we choose our target as seven-dimensional Lovelock AdS black holes. Lovelock gravity~\cite{Lovelock} successfully solved the problems of ghosts \cite{Boulware} and field equations of fourth order. So it is certainly of interest to probe the black hole solutions and their thermodynamics within the framework of Lovelock gravity~\cite{Dehghani1}-\cite{Amirabi}. Ref.~\cite{Dehghani1} found the topological black hole solutions in Lovelock-Born-Infeld gravity. Their thermodynamics and critical behavior were investigated in detail~\cite{Decheng4}-\cite{Dolan99}. In this paper, we would like to generalize the current research of Joule-Thomson expansion to the black hole solutions found in Ref.~\cite{Dehghani1}. This generalization is of physical significance since varieties of intriguing thermodynamic properties have been disclosed for this solution. It is naturally expected that the research in this paper may give rise to novel findings concerning the Joule-Thomson expansion. To concentrate on the effect of Lovelock gravity, we do not consider the nonlinear electromagnetic filed here.

    The organization of this paper is as follows. Sec.\ref{Sec2} devotes to a short review of the thermodynamics of $(n+1)$-dimensional Lovelock AdS black holes. In Sec.\ref {Sec3}, we will probe the Joule-Thomson expansion of seven-dimensional Lovelock AdS black holes. Conclusion will be presented in Sec.\ref {Sec4}.

\section{A brief review of the thermodynamics of $(n+1)$-dimensional Lovelock AdS black holes}

\label{Sec2}
The action of Lovelock gravity coupled with Born-Infeld electromagnetic filed can be written as \cite{Dehghani1}
\begin{equation}
I_{G}=\frac{1}{16\pi}\int d^{n+1}x\sqrt{-g}\big(-2\Lambda+\mathcal{L}_1+\alpha_2\mathcal{L}_2+\alpha_3\mathcal{L}_3+L(F)\big),\label{1}
\end{equation}%
 where
\begin{eqnarray}
\mathcal{L}_1&=&R,\label{2}
\\
\mathcal{L}_2&=&R_{\mu\nu\gamma\delta}R^{\mu\nu\gamma\delta}-4R_{\mu\nu}R^{\mu\nu}+R^2, \label{3}
\\
\mathcal{L}_3&=&2R^{\mu\nu\sigma\kappa}R_{\sigma\kappa\rho\tau}R^{\rho\tau}_{\;\;\;\;\mu\nu}+8R^{\mu\nu}_{\;\;\;\;\sigma\rho}R^{\sigma\kappa}_{\;\;\;\;\nu\tau}R^{\rho\tau}_{\;\;\;\;\mu\kappa}+24R^{\mu\nu\sigma\kappa}R_{\sigma\kappa\nu\rho}R^{\rho}_{\;\;\mu} \nonumber
\\
&\,&+3RR^{\mu\nu\sigma\kappa}R_{\sigma\kappa\mu\nu}+24R^{\mu\nu\sigma\kappa}R_{\sigma\mu}R_{\kappa\nu}+16R^{\mu\nu}R_{\nu\sigma}R^{\sigma}_{\;\;\mu}
\nonumber
\\
&\,&-12RR^{\mu\nu}R_{\mu\nu}+R^3,\label{4}
\\
L(F)&=&4\beta^2\left(1-\sqrt{1+\frac{F^2}{2\beta^2}}\right). \label{5}
\end{eqnarray}%
$\mathcal{L}_1$, $\mathcal{L}_2$, $\mathcal{L}_3$ and $L(F)$ represent the Einstein-Hilbert Lagrangian, Gauss-Bonnet Lagrangian, the third order Lovelock Lagrangian and Born-Infeld Lagrangian respectively with $\alpha_2$, $\alpha_3$ denoting the second and third order Lovelock coefficients and $\beta$ denoting the Born-Infeld parameter.

Ref.~\cite{Dehghani1} derived its $(n+1)$-dimensional black hole solution as
\begin{equation}
ds^2=-f(r)dt^2+\frac{dr^2}{f(r)}+r^2d\Omega^2, \label{6}
\end{equation}%
where
\begin{eqnarray}
f(r)&=&k+\frac{r^2}{\alpha}[1-g(r)^{1/3}],\label{7}\\
g(r)&=&1+\frac{3\alpha m}{r^n}-\frac{12\alpha \beta^2}{n(n-1)}\Big[1-\sqrt{1+\eta}-\frac{\Lambda}{2\beta^2}+\frac{(n-1)\eta}{(n-2)}\digamma(\eta)\Big].\label{8}
\end{eqnarray}%
$\digamma(\eta)$ denotes the hypergeometric function $\,_2F_1\Big(\Big[\frac{1}{2},\frac{n-2}{2n-2}\Big],\Big[\frac{3n-4}{2n-2}\Big],-\eta\Big)$ with $\eta=\frac{(n-1)(n-2)q^2}{2\beta^2r^{2n-2}}$.
It should be noted that this solution was obtained with the assumption that $\alpha_2=\frac{\alpha}{(n-2)(n-3)},\\ \alpha_3=\frac{\alpha^2}{72{n-2\choose 4}}$, where $\alpha$ denotes the Lovelock parameter.

In the above solution, $m$ and $q$ are parameters related to the mass $M$ and the electric charge $Q$ as
\begin{eqnarray}
M&=&\frac{(n-1)\Sigma_k}{16\pi}m,\label{9}
\\
Q&=&\frac{\Sigma_k}{4\pi}\sqrt{\frac{(n-1)(n-2)}{2}}q.\label{10}
\end{eqnarray}%
Here, $\Sigma_k$ denotes the volume of the $(n-1)$-dimensional hypersurface with constant curvature $(n-1)(n-2)k$ whose line element is represented by $d\Omega^2$ in the above metric. Note that the horizon radius $r_+$ is the largest root of the equation $f(r_+)=0$. Then the mass can be expressed into the function of $r_+$.

Since we need to concentrate on the effect of Lovelock gravity, the effect of nonlinear electromagnetic filed can be excluded by considering the limit $\beta\rightarrow\infty$. Then
\begin{equation}
g(r)\rightarrow1+\frac{3\alpha m}{r^n}+\frac{6\alpha \Lambda}{n(n-1)}-\frac{3\alpha q^2}{r^{2n-2}}.\label{11}
\end{equation}%
The mass, the Hawking temperature, the entropy and the electric potential $\Phi$ of topological Lovelock AdS black holes was derived as~\cite{Dehghani1}
\begin{eqnarray}
M&=&\frac{\Sigma_k}{48n\pi r_+^{n+6}}\{3n(n-1)q^2r_+^8+r_+^{2n}\left[kn(n-1)(3r_+^4+3kr_+^2\alpha+k^2\alpha^2)-6r_+^6\Lambda\right]\}.\nonumber \\
\label{12} \\
T&=&\frac{1}{12\pi(n-1)r_+(r_+^2+k\alpha)^2}\{-6\Lambda r_+^6-3(n-2)(n-1)q^2r_+^{8-2n}\nonumber
\\
&\,&+(n-1)k[3(n-2)r_+^4+3(n-4)k\alpha r_+^2+(n-6)k^2\alpha^2]\}.\label{13}
\\
S&=&\int^{r_+}_{0}\frac{1}{T}\left(\frac{\partial M}{\partial r_+}\right)dr=\frac{\Sigma_k(n-1)r_+^{n-5}}{4}\left(\frac{r_+^4}{n-1}+\frac{2kr_+^2\alpha}{n-3}+\frac{k^2\alpha^2}{n-5}\right),\nonumber
\\
\label{14}
\\
\Phi&=&\frac{4\pi Q}{(n-2)r_+^{n-2}\Sigma_k}.\label{15}
\end{eqnarray}%
The integration in Eq. (\ref{14}) is divergent for $n\leqslant5$. So only Lovelock AdS black holes in the seven-dimensional (or above) space-time  makes sense physically considering this constraint.

Treating the cosmological constant as a variable, the thermodynamic pressure and thermodynamic volume can be defined as
\begin{equation}
P=-\frac{\Lambda}{8\pi},\;\;\;V=\left(\frac{\partial M}{\partial P}\right)=\frac{r_+^{n}\Sigma_k}{n}.\label{16}
\end{equation}%

\section{Joule-Thomson expansion of seven-dimensional Lovelock AdS black holes}
\label{Sec3}

\subsection{The Joule-Thomson coefficient}

  As a criteria to discriminate between heating process and cooling process in the Joule-Thomson expansion, the Joule-Thomson coefficient attracts much attention. Its definition is as follow
\begin{equation}
\mu=\left(\frac{\partial T}{\partial P}\right)_{H}\;. \label{17}
\end{equation}
Since the pressure decreases during the expansion, $\mu>0$ corresponds to the cooling process occurs while $\mu<0$ corresponds to heating.

Now we utilize the first law of black hole thermodynamics to derive the Joule-Thomson coefficient for Lovelock AdS black holes. The first law of black hole thermodynamics of seven-dimensional Lovelock AdS black holes in the extended phase space reads \cite{jiexiong2,Kastor2}
\begin{equation}
dH=TdS+\Phi dQ+V dP+\mathcal {A}d\alpha,\label{18}
\end{equation}%
where $\mathcal {A}$ denotes the quantities conjugated to the Lovelock parameter. Note that the mass of the black hole is interpreted as the enthalpy in the extended phase space and the term $\mathcal{B}d\beta$ is omitted here because the nonlinear electromagnetic field is not taken into account here.

From Eq. (\ref{18}), one can obtain
\begin{equation}
0=T\left(\frac{\partial S}{\partial P}\right)_{H}+V\;. \label{19}
\end{equation}
The conditions $dH=0, dQ=0, d\alpha=0$ have been used in the above derivation. On the other hand, the entropy $S$ can be viewed as the function of the Hawking temperature $T$, the pressure $P$, the charge $Q$ and the Lovelock parameter $\alpha$. And one can express $dS$ as
\begin{equation}
dS=\left(\frac{\partial S}{\partial P}\right)_{T,\alpha,Q}dP+\left(\frac{\partial S}{\partial T}\right)_{P,\alpha,Q}dT+\left(\frac{\partial S}{\partial \alpha}\right)_{T,P,Q}d\alpha+\left(\frac{\partial S}{\partial Q}\right)_{T,P,\alpha}dQ\;, \label{20}
\end{equation}
from which one can further derive
\begin{equation}
\left(\frac{\partial S}{\partial P}\right)_{H}=\left(\frac{\partial S}{\partial P}\right)_{T,\alpha,Q}+\left(\frac{\partial S}{\partial T}\right)_{P,\alpha,Q}\left(\frac{\partial T}{\partial P}\right)_{H}\;. \label{21}
\end{equation}
Here, we have utilized $dQ=0, d\alpha=0$ again.

Substituting Eq. (\ref{17}) into (\ref{21}), one can obtain
\begin{equation}
\mu=\frac{1}{T\left(\frac{\partial S}{\partial T}\right)_{P,\alpha}}\left[-T\left(\frac{\partial S}{\partial P}\right)_{T,\alpha,Q}-V\right]=\frac{1}{C_{P,\alpha}}\left[T\left(\frac{\partial V}{\partial T}\right)_{P,\alpha,Q}-V\right]\;. \label{22}
\end{equation}
Note that both the Maxwell relation and the definition of specific heat have been applied.

One can also follow the approach proposed in Ref. \cite{Aydiner2} and derive the Joule-Thomson coefficient utilizing both the first law of black hole thermodynamics and the differentiation of the Smarr formula. We have shown in Ref. \cite{jiexiong3} both approaches are consistent with each other.

\begin{figure}[H]
\centerline{\subfigure[]{\label{1a}
\includegraphics[width=8cm,height=6cm]{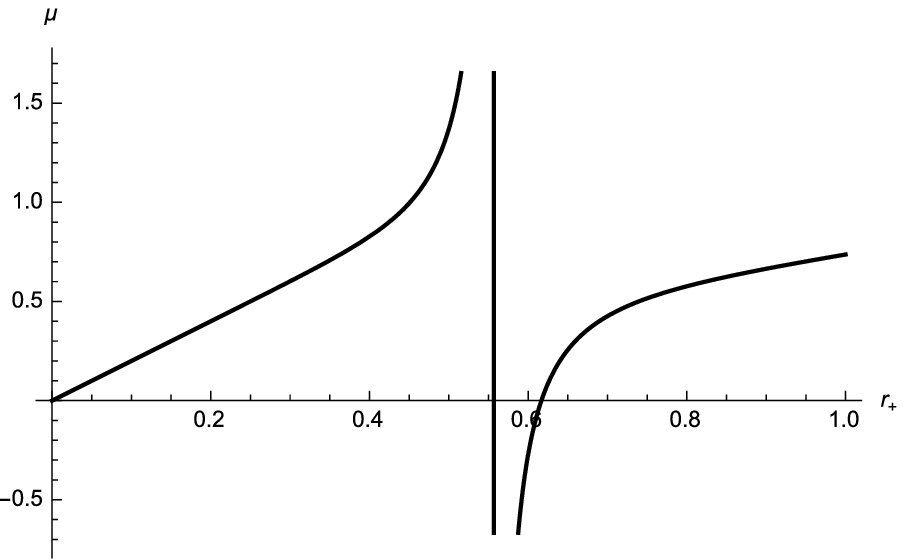}}
\subfigure[]{\label{1b}
\includegraphics[width=8cm,height=6cm]{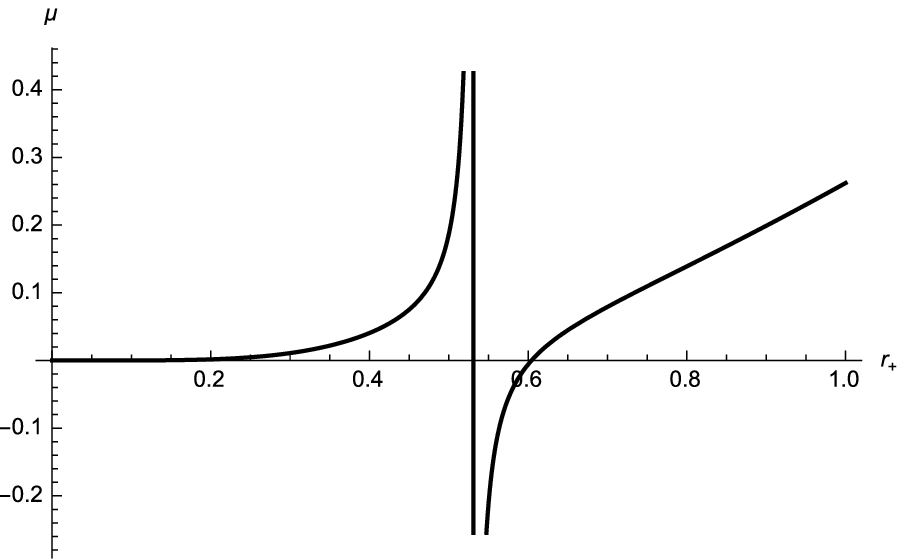}}}
\centerline{\subfigure[]{\label{1c}
\includegraphics[width=8cm,height=6cm]{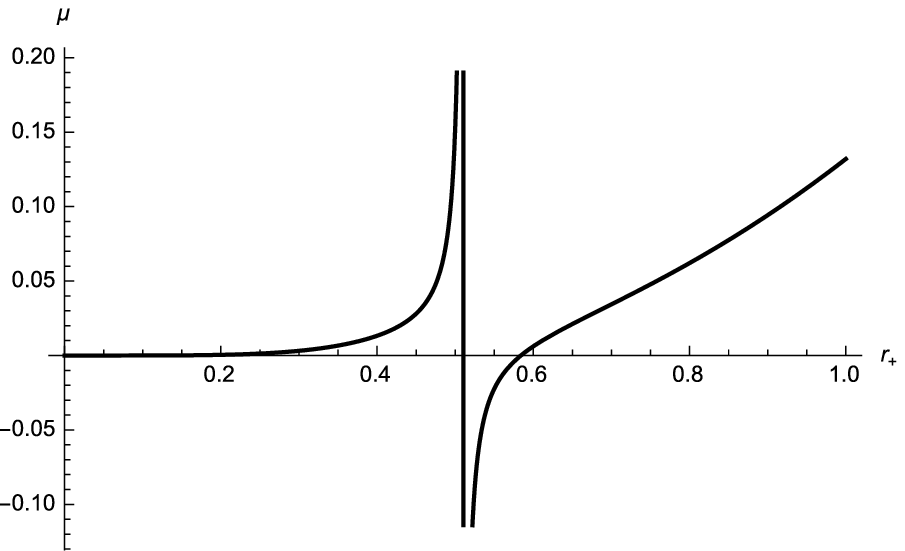}}
\subfigure[]{\label{1d}
\includegraphics[width=8cm,height=6cm]{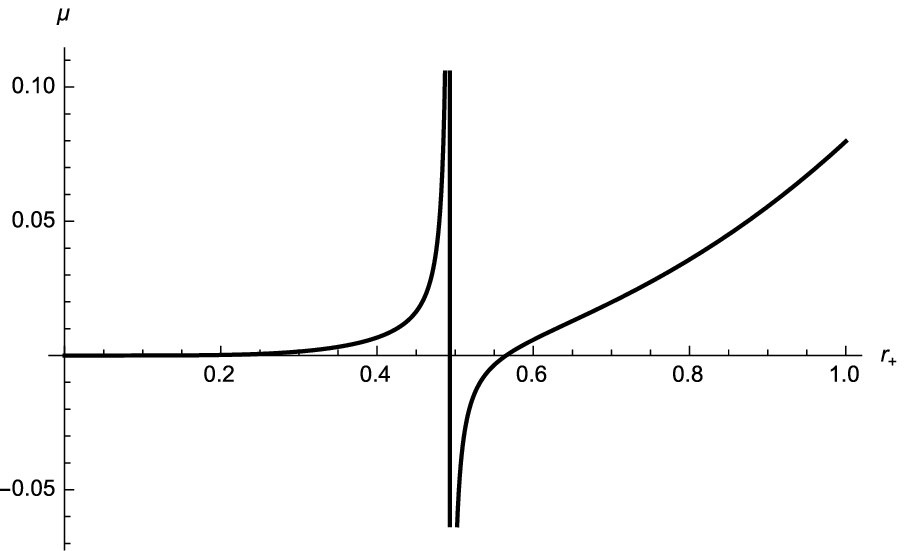}}}
\centerline{\subfigure[]{\label{1e}
\includegraphics[width=8cm,height=6cm]{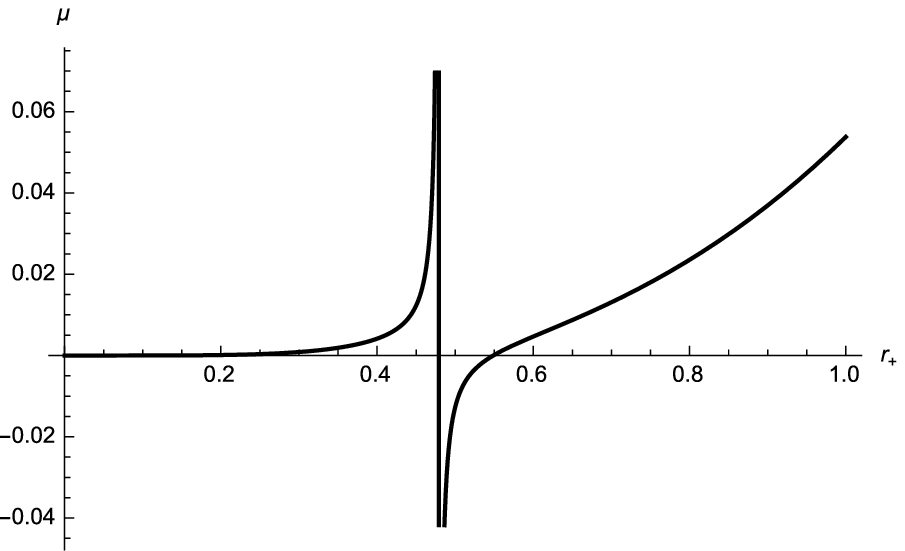}}
\subfigure[]{\label{1f}
\includegraphics[width=8cm,height=6cm]{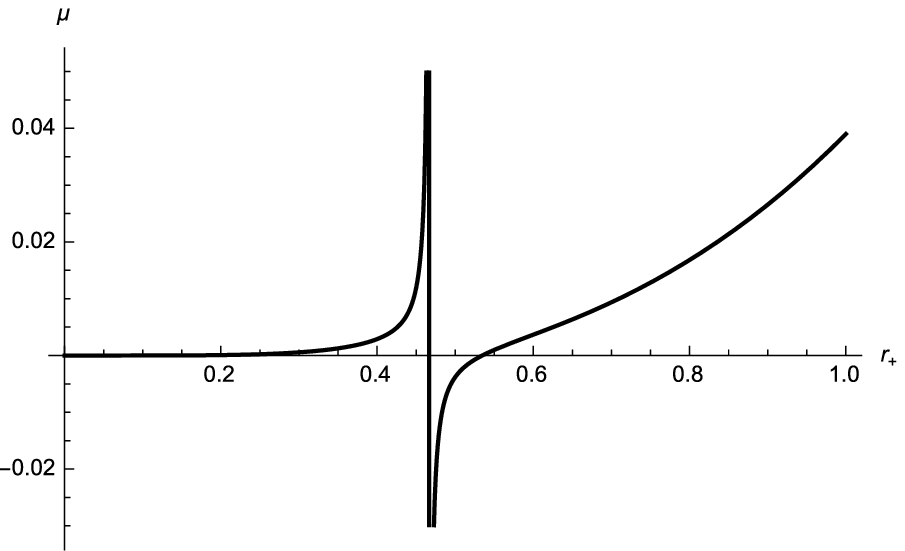}}}
 \caption{Joule-Thomson coefficient $\mu$ for $P=1, Q=1$ (a) $\alpha=0$ (b) $\alpha=0.5$ (c) $\alpha=1.0$ (d) $\alpha=1.5$ (e) $\alpha=2.0$ (f) $\alpha=2.5$}
\label{fg1}
\end{figure}

From Eqs. (\ref{13}), (\ref{14}), (\ref{16}) and (\ref{22}), the explicit expression of $\mu$ for seven-dimensional Lovelock AdS black holes can be obtained as

\begin{equation}
\mu=\frac{2r_+^5\{r_+^6[40P\pi r_+^6+75r_+^2\alpha+25\alpha^2+r_+^4(70+8P\pi \alpha)]-10q^2(15r_+^2+11\alpha)\}}{15(r_+^2+\alpha)^3(10r_+^8+8P\pi r_+^{10}+5\alpha r_+^6-10q^2)}\;. \label{23}
\end{equation}
The behavior of $\mu$ is depicted in Fig.\ref{fg1} with the help of Eq. (\ref{23}). Both a divergent point and a zero point can be observed in each subgraph. The horizon radius corresponding to the zero point (We use the notation $r_{+i}$ to denote this horizon radius while the subscript $i$ is short for ``inversion'' in this paper) tends to decrease with the Lovelock parameter $\alpha$. So does the horizon radius corresponding to the divergent point of $\mu$. Remind that the divergent point of Joule-Thomson coefficient coincides with the zero point of Hawking temperature.

Utilizing Eq. (\ref{13}), one can further obtain the inversion temperature $T_i$ when $P$, $Q$, $\alpha$ is given. Inversion temperature is an important quantity to probe the Joule-Thomson expansion. In the following subsection, we will study the inversion curves which show the relation between the inversion temperature $T_i$ and the inversion pressure $P_i$ intuitively.

\subsection{The inversion curves and the isenthalpic curves}

\begin{figure}[H]
\centerline{\subfigure[]{\label{2a}
\includegraphics[width=8cm,height=6cm]{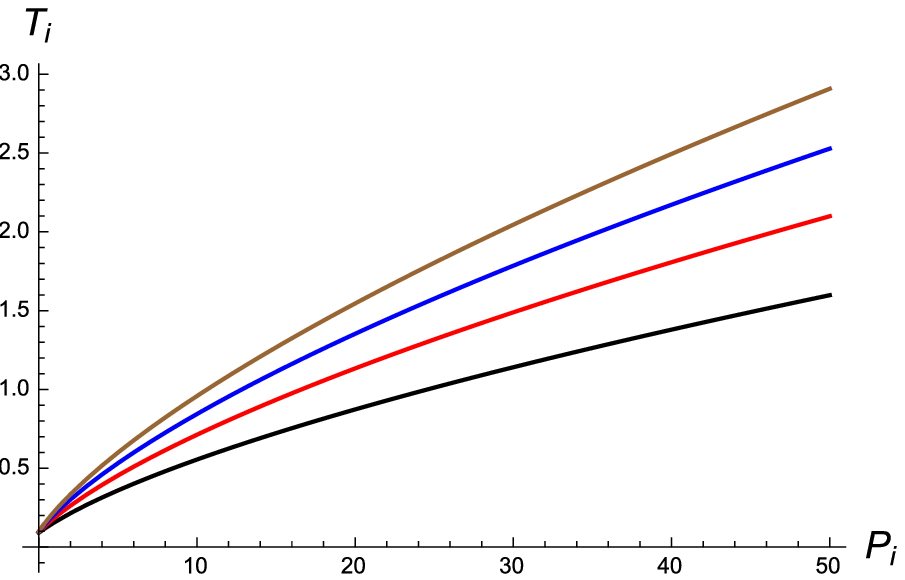}}
\subfigure[]{\label{2b}
\includegraphics[width=8cm,height=6cm]{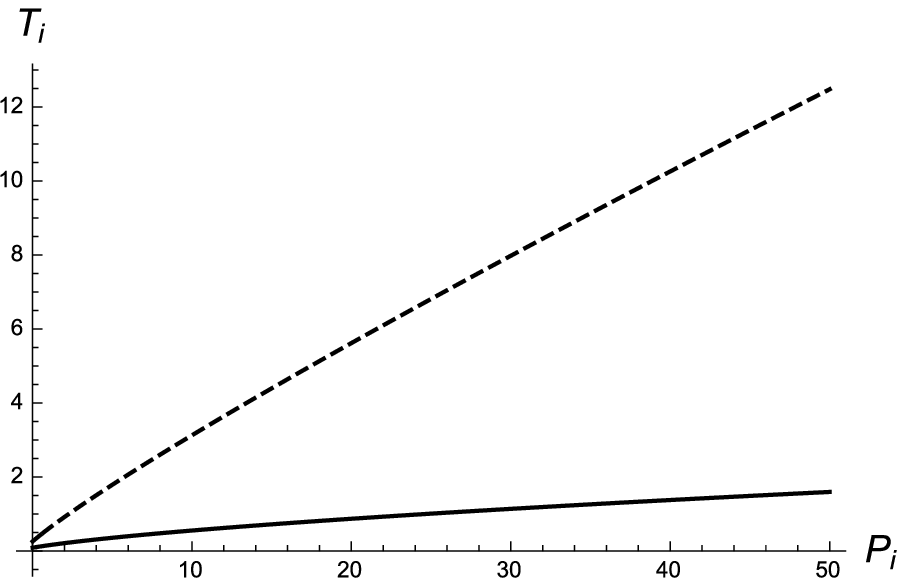}}}
\centerline{\subfigure[]{\label{2c}
\includegraphics[width=8cm,height=6cm]{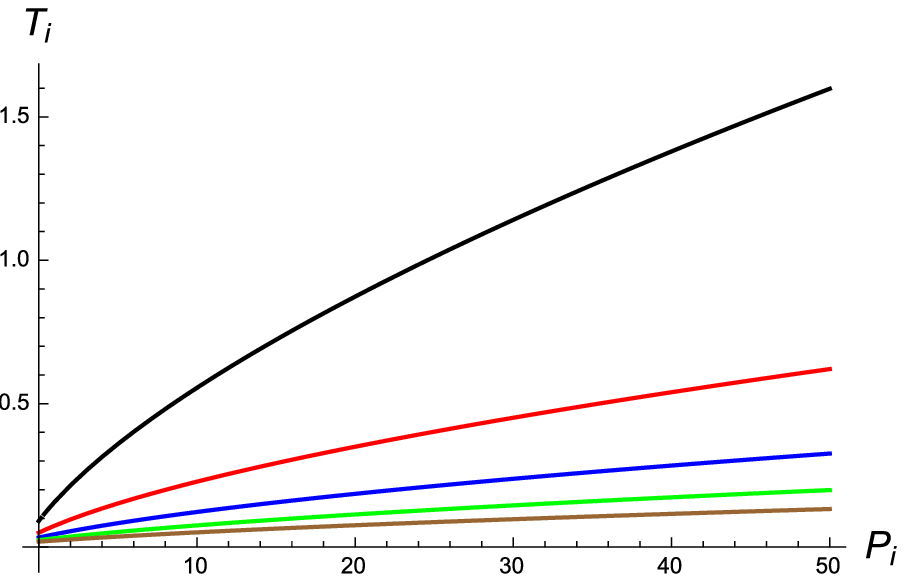}}
\subfigure[]{\label{2d}
\includegraphics[width=8cm,height=6cm]{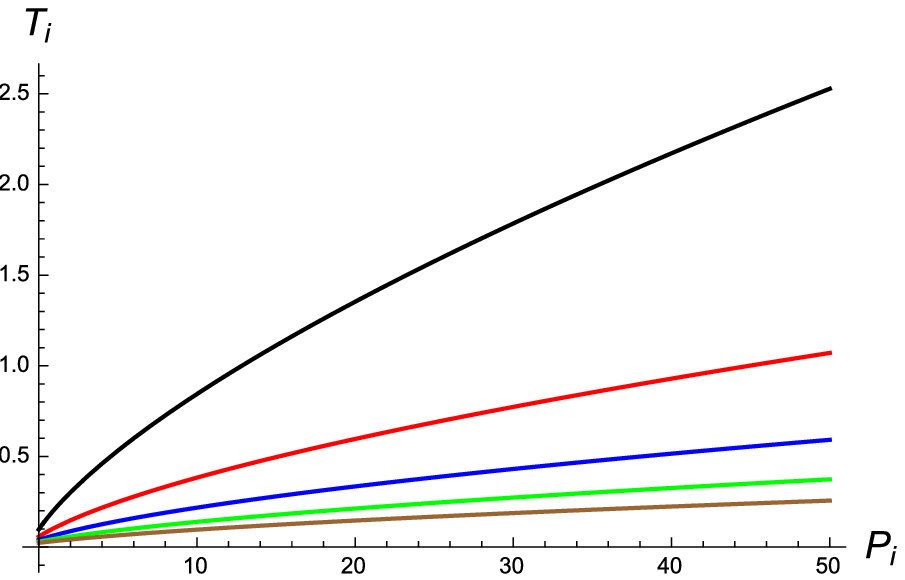}}}
 \caption{ (a)The inversion curves for $\alpha=0.5$. From top to bottom the curves correspond to $Q=2.5$,  $Q=2.0$, $Q=1.5$, $Q=1.0$ respectively.(b) The inversion curves for $Q=1$. The dashed line represents $\alpha=0$ while the solid line represents $\alpha=0.5$   (c) The inversion curves for $Q=1$. From top to bottom the curves correspond to $\alpha=0.5$,  $\alpha=1.0$, $\alpha=1.5$, $\alpha=2.0$, $\alpha=2.5$ respectively. (d) The inversion curves for $Q=2$. From top to bottom the curves correspond to $\alpha=0.5$,  $\alpha=1.0$, $\alpha=1.5$, $\alpha=2.0$, $\alpha=2.5$ respectively }
\label{fg2}
\end{figure}

As can be observed from the numerator of the righthand side of Eq. (\ref{23}), the inversion pressure $P_i$ and the corresponding $r_{+i}$ satisfy
 \begin{equation}
r_{+i}^6[40P_i\pi r_{+i}^6+75r_{+i}^2\alpha+25\alpha^2+r_{+i}^4(70+8P_i\pi \alpha)]-10q^2(15r_{+i}^2+11\alpha)=0\;. \label{24}
\end{equation}
Via Eqs. (\ref{13}), (\ref{16}) and (\ref{24}), the inversion curves are plotted in Fig.\ref{fg2}. For $\alpha=0.5$, Fig.\ref{2a} shows the effect of the electric charge on the inversion curves. With the increasing of $Q$, the inversion temperature for given pressure tends to increase. This behavior is qualitatively similar to the RN-AdS black holes \cite{Aydiner1}. Fig.\ref{2b} further compares the inversion curve of the case $\alpha=0.5$ with that of the case $\alpha=0$. Although in both case the inversion temperature $T_i$ increases with the inversion pressure $P_i$, the slope is rather different. For given $P_i$, the inversion temperature of the case $\alpha=0.5$ is much lower. To further probe the effect of the Lovelock gravity on the inversion curves, Fig.\ref{2c} and \ref{2d} shows the inversion curves for different $\alpha$ while $Q$ is chosen to be $1$ and $2$ respectively. As can be witnessed in these two sub-figures, $T_i$ tends to decrease with $\alpha$ for given $P_i$, in contrast to the effect of the electric charge.

Since $T_i$ increases monotonically with $P_i$, there exists a minimum inversion temperature $T_{min}$ which can be calculated by demanding $P_i=0$. The ratio between $T_{min}$ and the critical temperature $T_c$ is of potential interest. It was shown that this ratio is $1/2$ for four-dimensional charged AdS black holes \cite{Aydiner1} and Kerr-AdS black holes \cite{Aydiner2} but tends to decrease with the dimensionality $d$ \cite{jiexiong3}. Note that $T_{min}$ can be obtained by demanding $P_i=0$. The results of the ratio $T_{min}/T_c$ for various $\alpha$ is shown in Table \ref{tb1}. For $\alpha=0$, it recovers the value of $d=7$ \cite{jiexiong3}. For $\alpha\neq0$, the ratio decreases with $\alpha$.

\begin{table}[!h]
\centering
\begin{tabular}{|c|c|c|c|c|c|c|}
  \hline
$\alpha$ & 0\;\; & 0.5\;\;  &1.0\;\;  &1.5\;\;  &2.0\;\;  &2.5 \\   \hline
    $T_{min}$ \; & 0.257949 \;\;  &0.090899 \;\;  &      0.050021\;\;  & 0.033123\;\;  &0.024259\;\;   &0.018915 \\
     $T_c$ \; & 0.587673 \;\;&0.200981 \;\;  &   0.142337\;\;  &0.116228\;\;  &0.100658\; \;  &0.090031 \\
      $T_{min}/T_c$ \; & 0.438933 \;\;\; &0.452277 \;\;\; &      0.351427\;\;\; & 0.284983\;\;\; &0.241004\;\;\;  &0.210094  \\ \hline
       \end{tabular}
       \caption{$T_{min}/T_c$ of seven-dimensional charged Lovelock AdS black holes for various $\alpha$ ($Q=1$)}
       \label{tb1}
       \end{table}

\begin{figure}[H]
\centerline{\subfigure[]{\label{3a}
\includegraphics[width=8cm,height=6cm]{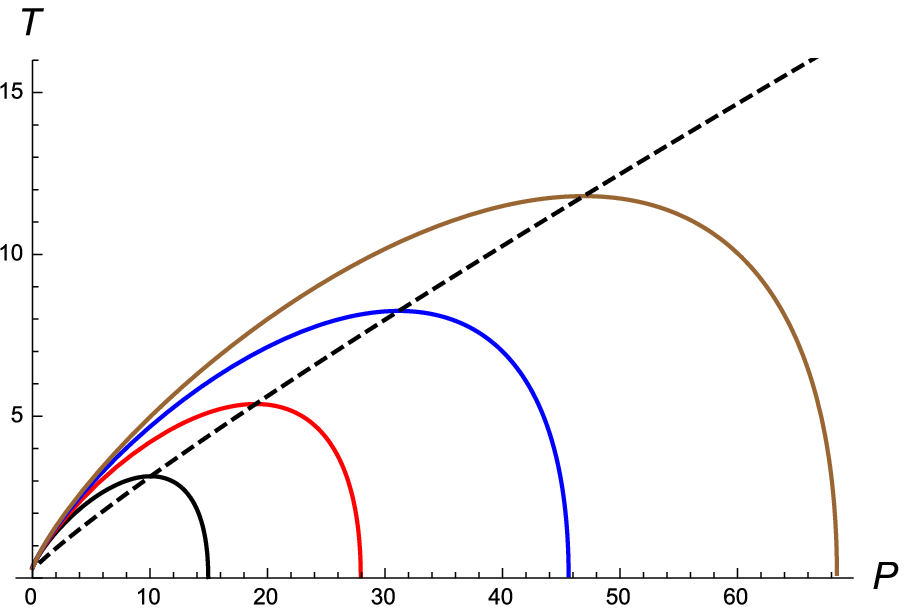}}
\subfigure[]{\label{3b}
\includegraphics[width=8cm,height=6cm]{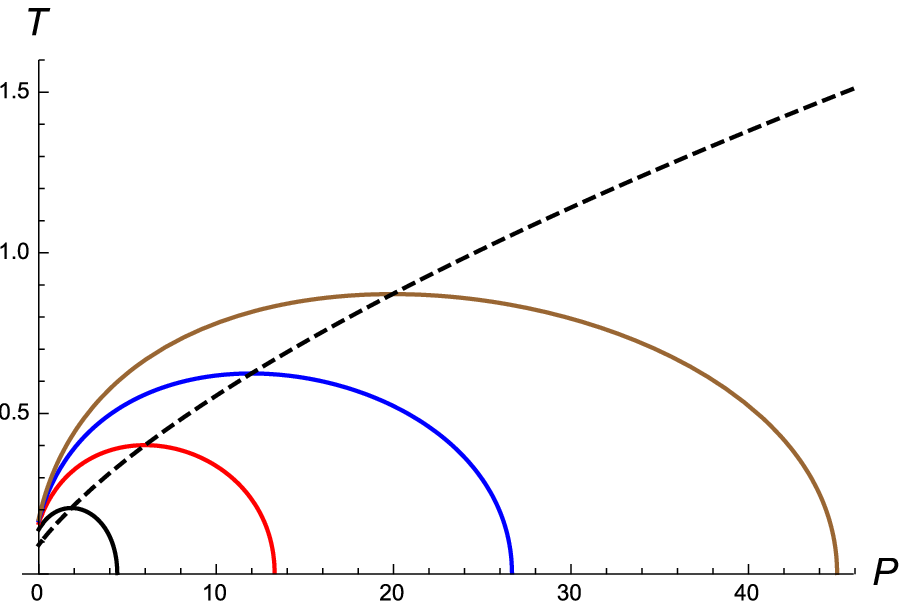}}}
 \caption{Isenthalpic curves for $Q=1$ (a) $\alpha=0$ (b) $\alpha\neq0$ ($\alpha=0.5$). In both graphs the curves from left to right correspond to $M=2.0$, $M=2.5$, $M=3.0$, $M=3.5$ respectively. Note that the inversion curve for $Q=1$ is also depicted in both graphs via the dashed line.}
\label{fg3}
\end{figure}

Apart from the inversion curves, the isenthalpic curves are also of interest considering Joule-Thomson expansion is an isenthalpic process. Within the framework of the extended phase space, the mass is interpreted as enthalpy \cite{Kastor}. Then isenthalpic curves can be plotted provided the mass of the black holes is fixed. Utilizing Eqs. (\ref{12}), (\ref{13}) and (\ref{16}), the isenthalpic curves for $\alpha=0$ and $\alpha\neq0$ is shown in Fig.\ref{fg3}. The inversion curves are shown together to gain a deeper understanding of the Joule-Thomson expansion. Each isenthalpic curve is divided into two branches by the inversion curve. The branch above the inversion curve represents the cooling process while the branch below the inversion curve represents the heating process. Note that the intersection point between the isenthalpic curve and the inversion curve is exactly the inversion point. So the inversion point discriminates the heating process from cooling process. Moreover, the inversion point is the maximum point for a specific isenthalpic curve, implying that during the whole Joule-Thomson expansion process the temperature is highest at the inversion point.

One can read off the inversion temperature and the inversion pressure from the intersection point between the isenthalpic curve and the inversion curve. Comparing the isenthalpic curves for $\alpha\neq0$ with those of the case $\alpha=0$, it can be witnessed that both $T_i$ and $P_i$ for $\alpha\neq0$ are much lower for the same given mass of the black hole, showing the effect of Lovelock gravity. Moreover, we list the corresponding $T_i$ and $P_i$ for various mass in Table \ref{tb2}. As $M$ increases, both $T_i$ and $P_i$ increase, suggesting that the inversion temperature is higher for the Joule-Thomson expansion with a larger enthalpy.

\begin{table}[!h]
\centering
\begin{tabular}{|c|c|c|c|}
  \hline
$\alpha$ & $M$\;\; & $P_i$\;\;  &$T_i$ \\   \hline
    0 \; & 2.0 \;\;  &9.99\;\;  &      3.13 \\
     0 \; & 2.5 \;\;&19.04 \;\;  &   5.39 \\
      0 \; & 3.0\;\;&31.17 \;\;  &   8.26 \\
       0 \; & 3.5 \;\;&47.02 \;\;  &   11.77 \\
        0.5 \; & 2.0 \;\;&1.70 \;\;  &   0.21 \\
          0.5 \; & 2.5 \;\;&5.95 \;\;  &   0.40 \\
            0.5 \; & 3.0 \;\;&11.94 \;\;  &   0.63\\
              0.5 \; & 3.5\;\;&19.87 \;\;  &   0.87 \\  \hline
       \end{tabular}
       \caption{Inversion temperature and inversion pressure of seven-dimensional charged Lovelock AdS black holes for various mass ($Q=1$)}
       \label{tb2}
       \end{table}

\subsection{An interesting case---uncharged Lovelock AdS black holes}

\begin{figure}
\centerline{\subfigure[]{\label{4a}
\includegraphics[width=8cm,height=6cm]{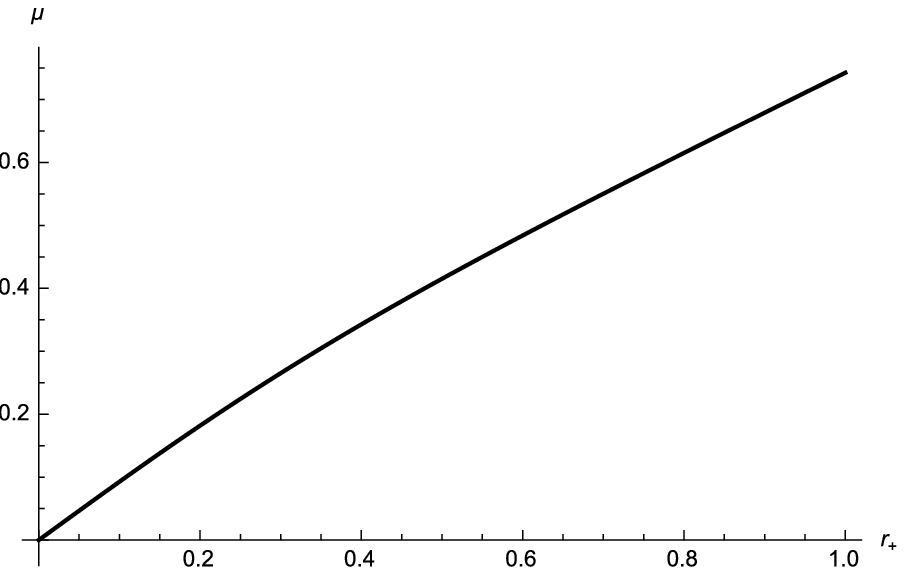}}
\subfigure[]{\label{4b}
\includegraphics[width=8cm,height=6cm]{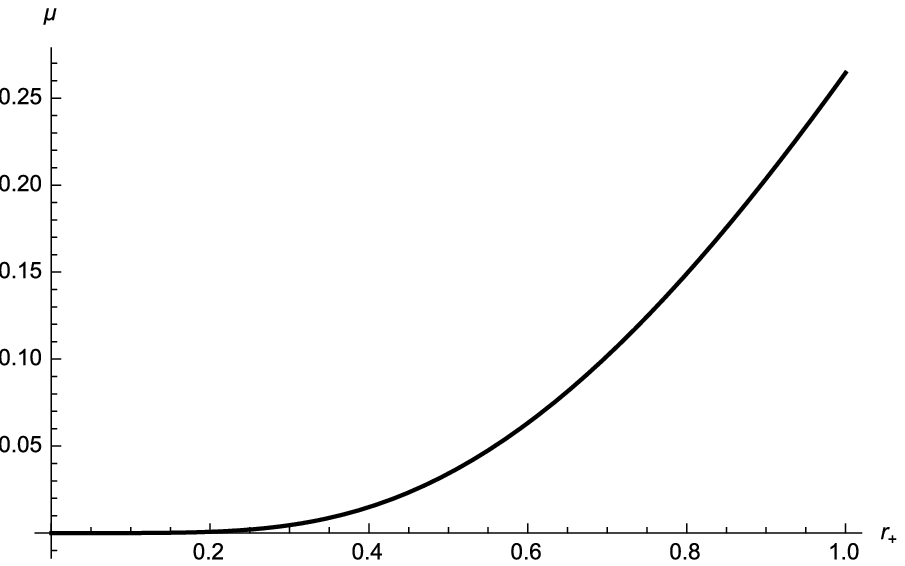}}}
\centerline{\subfigure[]{\label{4c}
\includegraphics[width=8cm,height=6cm]{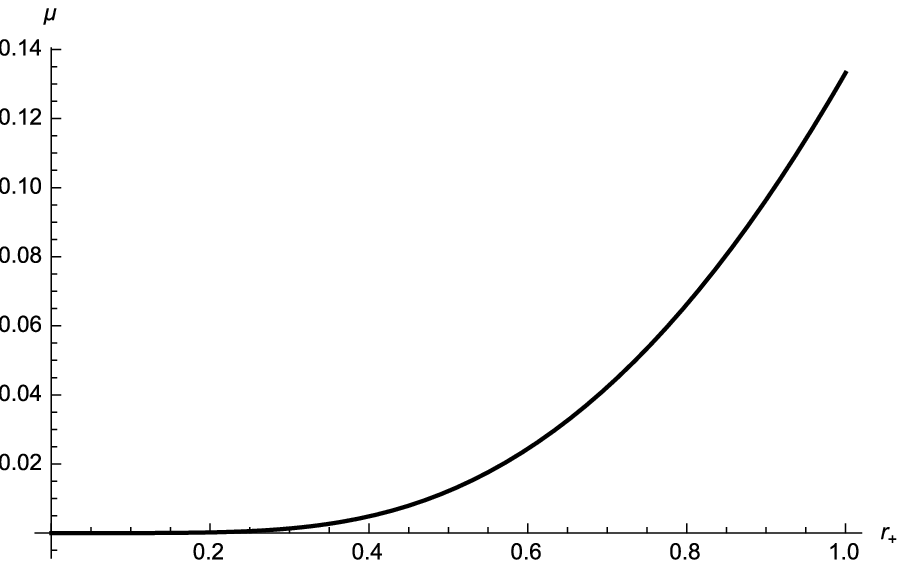}}
\subfigure[]{\label{4d}
\includegraphics[width=8cm,height=6cm]{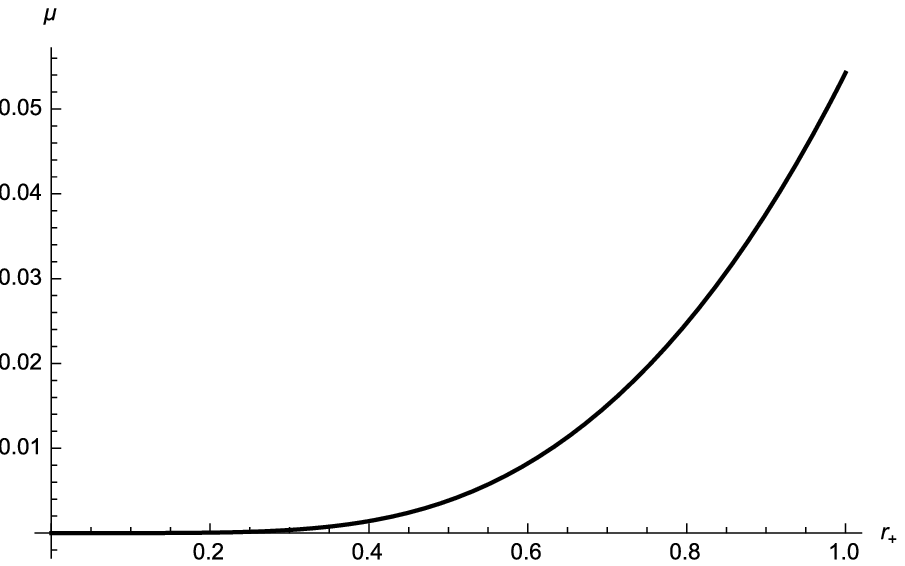}}}
 \caption{Joule-Thomson coefficient $\mu$ for $P=1, Q=0$ (a) $\alpha=0$ (b) $\alpha=0.5$ (c) $\alpha=1.0$ (d) $\alpha=2.0$}
\label{fg4}
\end{figure}

\begin{figure}[H]
\centerline{\subfigure[]{\label{5a}
\includegraphics[width=8cm,height=6cm]{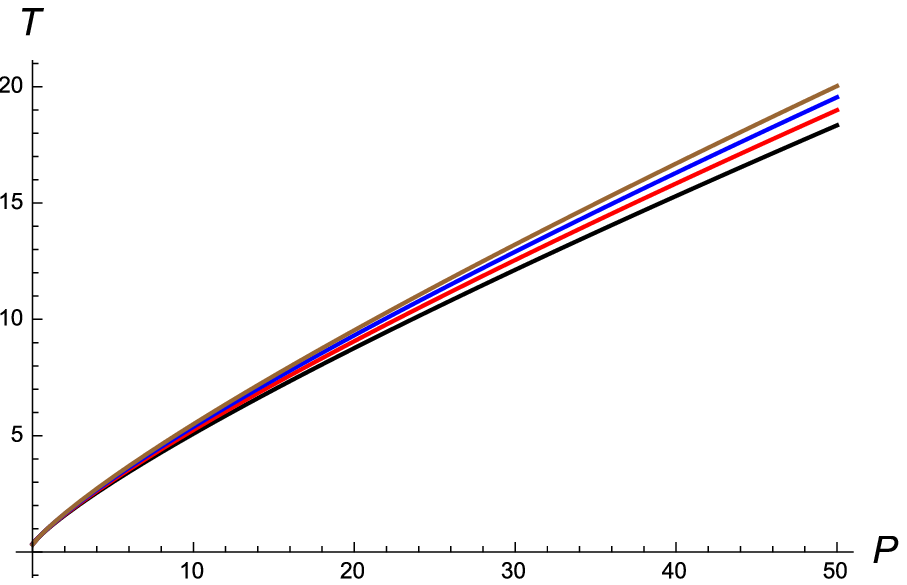}}
\subfigure[]{\label{5b}
\includegraphics[width=8cm,height=6cm]{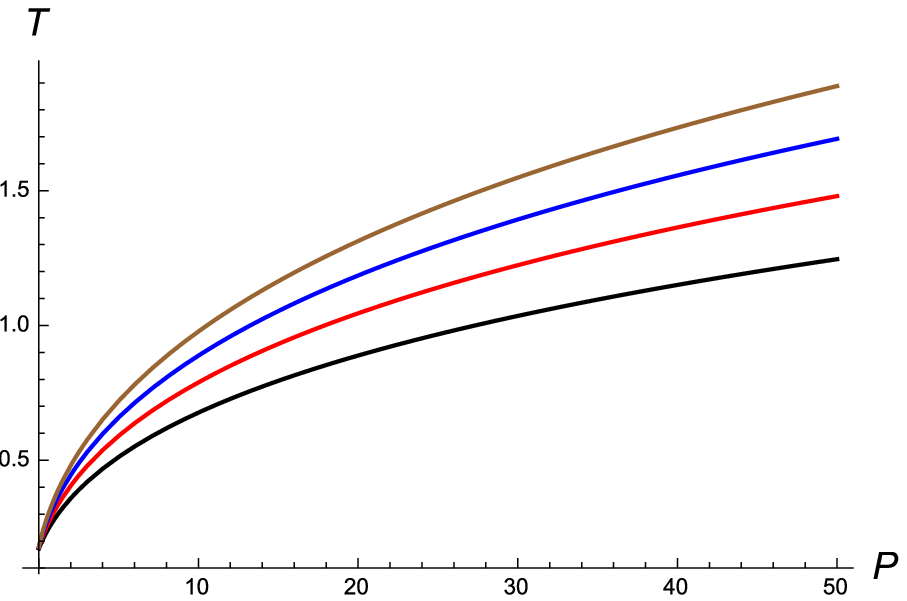}}}
 \caption{Isenthalpic curves for $Q=0$ (a) $\alpha=0$ (b) $\alpha\neq0$ ($\alpha=0.5$). In both graphs the curves from bottom to top correspond to $M=2.0$, $M=2.5$, $M=3.0$, $M=3.5$ respectively. }
\label{fg5}
\end{figure}

Here, we consider an interesting case---the uncharged Lovelock AdS black holes. From Eqs. (\ref{13}), (\ref{14}), (\ref{16}) and (\ref{22}), one can obtain the explicit expression of $\mu$ as
\begin{equation}
\mu=\frac{2r_+^5[40P\pi r_+^6+75r_+^2\alpha+25\alpha^2+r_+^4(70+8P\pi \alpha)]}{15(r_+^2+\alpha)^3(10r_+^2+8P\pi r_+^4+5\alpha)}\;. \label{25}
\end{equation}
As can be observed from above, both its numerator and denominator are always positive since $P>0, \alpha\geq0$. So there will be no zero point or divergent point for $\mu$. This conclusion can also be confirmed in Fig.\ref{fg4}, where the behavior of $\mu$ is plotted for the case $Q=0$ for various choices of $\alpha$. $\mu$ is always positive, suggesting the expansion is always in the regime of cooling process. Then no inversion temperature exists in this case, in contrast to the case $Q\neq0$. Isenthalpic curves are presented in Fig.\ref{fg5} for both the case $\alpha=0$ and $\alpha\neq0$. It is shown in both cases that the temperature increases monotonically with the pressure when the mass is specified. This observation is also quite different from that in the isenthalpic curves of $Q\neq0$.

\section{Conclusions}
\label{Sec4}
Effects of Lovelock gravity on the Joule-Thomson expansion is probed in this paper from various perspectives. Specifically, we investigate the Joule-Thomson expansion of seven-dimensional Lovelock AdS black holes.

Firstly, the well-known Joule-Thomson coefficient $\mu$ is derived via the first law of black hole thermodynamics. The explicit expression of $\mu$ is obtained while its behavior is also shown intuitively. Both a divergent point and a zero point exist. Here, the divergent point corresponds to the zero point of Hawking temperature while the zero point is the inversion point that discriminate the cooling process from heating process. It is shown that the horizon radius corresponding to the zero point tends to decrease with the Lovelock parameter $\alpha$. So does the horizon radius corresponding to the divergent point of $\mu$.

Secondly, the inversion curves showing the relation between the inversion temperature and the inversion pressure are studied. With the increasing of $Q$, the inversion temperature for given pressure tends to increase. This behavior is qualitatively similar to the RN-AdS black holes. Comparing the inversion curve of the case $\alpha\neq0$ with that of the case $\alpha=0$, it is shown that in both case the inversion temperature $T_i$ increases with the inversion pressure $P_i$. However, the slope is rather different. For given $P_i$, the inversion temperature of the case $\alpha\neq0$ is much lower. Moreover, the inversion temperature tends to decrease with $\alpha$ for given $P_i$, in contrast to the effect of the electric charge. We also probe the ratio between $T_{min}$ and the critical temperature $T_c$. It is shown that the ratio $T_{min}/T_c$ decreases with $\alpha$ for $\alpha\neq0$ while it recovers the result in former literature \cite{jiexiong3} for $\alpha=0$.

 Thirdly, the isenthalpic curves are investigated with rich physics revealed. Each isenthalpic curve is divided into two branches by the inversion curve. The branch above the inversion curve represents the cooling process while the branch below the inversion curve represents the heating process. And the intersection point between the isenthalpic curve and the inversion curve is exactly the inversion point. So the inversion point discriminates the heating process from cooling process. Moreover, the inversion point is the maximum point for a specific isenthalpic curve, implying that during the whole Joule-Thomson expansion process the temperature is highest at the inversion point. One can read off the inversion temperature and the inversion pressure from the intersection point between the isenthalpic curve and the inversion curve. Comparing the case $\alpha\neq0$ with the case $\alpha=0$, it can be witnessed that both $T_i$ and $P_i$ for $\alpha\neq0$ are much lower for the same given mass of the black hole, showing the effect of Lovelock gravity. Moreover, both $T_i$ and $P_i$ increase as $M$ increases, suggesting that the inversion temperature is higher for the Joule-Thomson expansion with a larger enthalpy.

Last but not the least, an interesting case---uncharged Lovelock AdS black holes is discussed. It is shown that $\mu$ is always positive, suggesting the expansion is always in the regime of cooling process. Then no inversion temperature exists in this case, in contrast to the case $Q\neq0$. Isenthalpic curves are also quite different from those of $Q\neq0$. It is shown for both the cases $\alpha\neq0$ and $\alpha=0$ that the temperature increases monotonically with the pressure when the mass is specified.

 \acknowledgments This research is supported by National Natural Science Foundation of China (Grant No.1160\\5082). The authors are also in part supported by National Natural Science Foundation of China (Grant No.11747017), Natural Science Foundation of Guangdong Province, China (Grant Nos.2016A030310363, 2016A030307051, 2015A030313789) and Department of Education of Guangdong Province, China (Grant Nos.2017KQNCX124,2017KZDXM056).

\end{document}